\documentclass[aps,prl,twocolumn,showpacs,superscriptaddress,groupedaddress]{revtex4}  
\usepackage{graphicx}  
\usepackage{epstopdf}
\usepackage{dcolumn}   
\usepackage{bm}        
\usepackage{amssymb}   
\usepackage{amsmath}   
\usepackage[version=4]{mhchem}
\usepackage[utf8]{inputenc}
\usepackage{textcomp}
\usepackage{float}
\usepackage{hyperref}
\usepackage{booktabs} 
\usepackage{siunitx}
\usepackage{newunicodechar}
\newunicodechar{₂}{$_2$}
\usepackage{afterpage}
\usepackage{tabularx,array}
\newcolumntype{L}{>{\raggedright\arraybackslash}X} 
\newcolumntype{C}[1]{>{\centering\arraybackslash}m{#1}} 
\usepackage[usenames, dvipsnames]{color}
\usepackage[section]{placeins} 

\begin{document}

\title{Nonadiabaticity under compression in metastable carbon monoxide-nitroxide mixtures}
\author{Reetam Paul} \affiliation{Lawrence Livermore National Laboratory, 7000 East Avenue, Livermore, CA 94550}
\author{Jonathan C. Crowhurst} \affiliation{Lawrence Livermore National Laboratory, 7000 East Avenue, Livermore, CA 94550}
\author{Stanimir A. Bonev} \affiliation{Lawrence Livermore National Laboratory, 7000 East Avenue, Livermore, CA 94550}


\begin{abstract}

Carbon monoxide (CO) and nitrous oxide (N₂O) both undergo profound structural and chemical transformations when compressed.  While their individual high–$P$/$T$ phase diagrams have been mapped in considerable detail, comparatively little attention has been paid to the \emph{mixtures} in which the two species can couple through oxygen transfer, charge redistribution, and nonadiabatic dissociation pathways. Here we use comprehensive \textit{ab initio} adiabatic/nonadiabatic molecular dynamics simulations, essentially a \emph{diabatic trajectory stitching} approach, that chart the evolution of CO--N$_2$O mixtures from van-der-Waals fluids to extended amorphous network solids over the range $0$--$160$~GPa and $300$--$1500$~K. We emphasize on (i) the sequence of gas\,$\rightarrow$\,molecular crystal\,$\rightarrow$\, polymerized amorphous solid reactive transitions that arise from an interplay between thermal and compression effects in metastable \ce{C-N-O} mixtures, (ii) the role of N$_2$O unimolecular dissociation in lowering the onset pressure for CO polymerization, and (iii) the emergence of nonadiabatic pathways, via thermal unimolecular dissociation of N$_2$O, accompanied by spin-transition in oxygen atoms that can make C-N-O systems deviate from Born-Oppenheimer dynamics. This dominates the chemistry once the mixture enters the regime of bond-breaking temperatures ($T\gtrsim\SI{900}{K}$). 

\end{abstract}

\maketitle

\begin{figure*}[!t]
\includegraphics[width=0.95\textwidth]{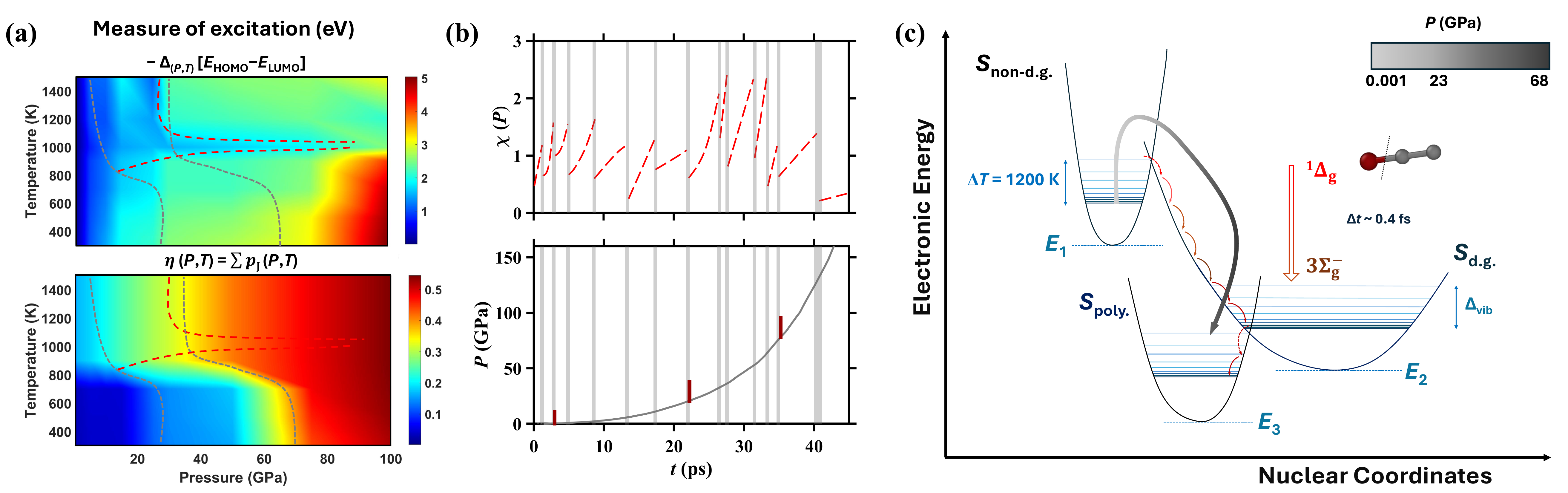} 
\caption{General characteristics of a C-N-O system, exemplified here with data from a CO-\ce{N2O} system: (a) The extent of excitation in a C-N-O system quantified as decrease in HOMO-LUMO gap as a function of pressure and temperature in the final configuration at each \textit{P-T} point. (b) The evolution of pressure with simulation time, along the 1200~K isotherm, highlighting in gray the specific time domains where NAMD has been performed. Red bars indicate phase transition boundaries. (c) A cartoon describing the nonadiabatic PES hopping (S$_{non-d.c.}$ to S$_{d.c.}$) as a result of singlet-oxygen dissociating from nitroxide molecules to form triplet-oxygen molecules; followed by equilibration with redistribution of energy ($\Delta_{vib}$) from vibrational to translational or rotational modes to either a dissociated state ($S_{d.c.}$) or a polymeric amorphous state ($S_{poly.}$).}
\label{fig:1}
\end{figure*}

Energetic materials store large amounts of chemical energy in the form of strained or highly polarized bonds \cite{zlotin2021advanced}, releasing a large fraction of that energy when those bonds relax to lower-energy configurations \cite{bernstein2014release,clemmer2023exploring}. One route to such states is the high-energy-density (HED) route of pressure--induced polymerization of small, stable molecules into extended networks in which every atom sits in an energetically \textit{over-bonded} environment \cite{eremets2008synthesis,lipp2005high,bonev2021energetics}. Because such transformations are driven by extreme compression and often high temperatures, the resulting frameworks can trap a substantial portion of the compression (\(P\,V\)) work done on them; if the phase can be metastably quenched and recovered to ambient conditions, that stored mechanical work becomes latent chemical energy. These high-pressure amorphous or disordered structures $-$ often lacking long-range crystalline order yet comprising robust covalent networks $-$ can potentially be used as energy-dense carbonaceous poly-nitro oxidizers \cite{kettner2016synthesis}. Two extensively researched examples illustrate this concept. Compressed carbon dioxide phase~V (\(\mathrm{CO_2}\)-V), which forms above 60~GPa \cite{yong2016crystal}, where molecular \(\mathrm{CO_2}\) polymerizes into a three-dimensional, four-coordinated, diamond-like sp\(^3\) C-O framework analogous to silica; on rapid decompression to ambient pressure it can survive below 185~K \cite{yong2016crystal}. Likewise, cubic gauche nitrogen (\textit{cg}-N) $-$ synthesized near 110~GPa and $>$2000~K $-$ locks every nitrogen into single, gauche-conformed N-N bonds, trapping the $\sim$781 kJ/mol enthalpic penalty of delocalizing the triple bonds of \(\mathrm{N_2}\) \cite{eremets2008synthesis}. Although both cases require diamond anvil cell conditions for synthesis and have prohibitive kinetic barriers and/or limited ambient recoverability \cite{yurtseven2016pippard,datchi2017polymeric}, they demonstrate how high-pressure amorphous or disordered solids could underwrite a new class of super-energetic materials once scalable routes to metastable recovery are developed.

This led us to investigate mixtures which are fundamentally C-N-O based but built using metastable mixtures of carbon monoxide (\ce{CO}) and nitroxides (in this case, nitrous oxide \ce{N2O}). The expectation was that these might be more energetic and recoverable relative to the baseline stable molecular cases of \ce{CO2} and \ce{N2}. Stoichiometrically, such a mixture is equivalent to an equimolar mixture of molecular \ce{CO2} and \ce{N2}. 

\ce{CO} polymerizes above $P\simeq\,5–10$~{GPa} to a Pa$\bar{3}$ cubic lattice at 300~K, forming layered sp$^{2}$$–$sp$^{3}$ carbon-oxygen frameworks with energy densities rivaling energetic polymers and certain nitramines. Above that, infrared and X-ray \cite{Evans2006ChemMater} signatures of $\pi$-bond collapse appear, signalling the formation of a lactone-type polymer \cite{Evans2005OSTI}; the product densifies smoothly to $\rho \approx \SI{2.6}{g\,cm^{-3}}$ by 40~GPa. Recent \textit{ab-initio} structure searches revealed even denser layered polymorphs (Cmcm, $P2_1/c$) that become more stable than previously proposed chains beyond 90~GPa \cite{Xia2017PRB,Sun2021PRB}. 

\ce{N2O}, in contrast, remains molecular till much higher pressures; remaining linear and centrosymmetric much further: Phase I (Pa$\bar{3}$) persists to 10 GPa \cite{Mills1991JCP}, Phase II (bent, $C2/m$) appears at $\sim\,$12 GPa \cite{Mills1991JCP}, and the fully bent \textit{Pbcn} Phase IV is accessed only above 18 GPa and 600 K \cite{Iota2004PRB}, before ultimately undergoing heterolytic disproportionation into NO$^{+}$NO$_3^{-}$ and N$_2$ \cite{Yoo2003JPCB}. The key divergence, at high pressures, from the CO$_2$ analogy is that bent \ce{N2O} eventually \emph{disproportionates} and does not \emph{polymerize}, once $P$ and $T$ are high enough to render the N$-$O bond heterolytic. This dominant thermal breakdown pathway is the spin-forbidden dissociation \cite{HerzbergDiatomic1950}
\[
\mathrm{N_2O}(\tilde{X}\,^1\Sigma^+) \;\longrightarrow\; \mathrm{N_2}(\tilde{X}\,^1\Sigma_g^+) + \mathrm{O}(^3\!P),
\]
which requires a change in spin-multiplicity and therefore proceeds via electronic-state crossing across potential energy surfaces (PES) rather than on a single adiabatic PES \cite{Hwang2000ChemPhys}. Non-adiabatic surface-hopping trajectories on high-level multireference PESs confirm that the crossing occurs at an extended region $\sim$1.8 Å along the N$–$O stretch, a geometry frequently sampled once $P\gtrsim1.5$ GPa and $T\gtrsim900$ K \cite{Subotnik2016ARPC}. 

In CO$–$N$_2$O mixtures, we can \emph{hypothesize} that this hot and highly rovibrationally excited O($^3\!P$) fragments can further oxidize nascent CO polymers; their exothermic recombination with developing C–O frameworks oxidizing three-coordinate carbon sites to CO$_2$. This would inject local heat, and generates point-defect disorder that frustrates any incipient crystallization, coupling unimolecular N$_2$O decomposition to the overall amorphization chemistry. We can speculate that this might drive down the amorphization pressures; with amorphous structures being preferable over crystalline ones for the sake of recoverability \cite{Santoro2006Nature}. Pressure‐driven polymerization of CO releases on the order of 3.5$–$4.0 eV per monomer \cite{Xia2017PRB,Sun2021PRB}, favoring network formation once intermolecular separations fall below $\approx$3 Å \cite{Evans2006ChemMater,Xia2017PRB}. Laser-pump experiments have shown that electronic excitations can lower the required pressure for network formation by $\approx$20\% \cite{Evans2006ChemMater}, emphasizing the role of non-adiabatic pathways even in the solid state. Thus, we can reason that the local heat released ($\approx$1 eV per reaction) and the spin-flip crossings along the N$-$O stretch can potentially act as an internal \emph{kinetic absorbers}, driving the system towards amorphous topologies even under quasi-isothermal external conditions. Overall, the outcome can potentially be a highly heterogeneous material in which densification, \ce{NO} radical-mediated oxidation, and nonadiabaticity–driven bond rearrangements occur on overlapping time scales ($\sim$0.1 to $\sim$1000~fs).

Because the \ce{N2O} decomposition involves a change in spin multiplicity, nuclei move on a \emph{manifold} of coupled electronic states.  A single adiabatic potential‐energy surface cannot describe the trajectory once the system approaches the crossing. Born–Oppenheimer molecular dynamics will therefore misrepresent both the timing and energetics of bond breaking.  Non-adiabatic trajectory‐surface-hopping \cite{Subotnik2016ARPC} or Ehrenfest dynamics \cite{Curchod2018ChemRev} — explicitly including spin–orbit coupling — is required to capture the competition between mechanical work done by compression (favoring polymerization) and electronic energy released by spin‐flipping N$_2$O cleavage.

This pursuit not only unlocks findings pertaining to the high-pressure physics and chemistry of such mixtures but also demonstrates methodological limitations of using \textit{ab initio} Born-Oppenheimer molecular dynamics in systems where oxygen atoms in nitroxides undergo a slow timescale (0.1-0.5 fs) spin-transition from singlet-to-triplet-to-singlet during the unimolecular thermal decomposition of \ce{N2O} that is concurrent with pressure-induced C-N-O amorphization; a process slow enough to be comparable to the time scales of ionic motion (0.5-1.0 fs). Such a sequence/overlap of transitions under combined pressure-temperature effects, which is fundamentally nonadiabatic in nature \cite{bhattacharya2010nonadiabatic} and being reported for the first time in the HED context, is the main crux of this letter.    

Thus, in this letter, we combine \emph{ab initio} Born-Oppenheimer adiabatic molecular dynamics (referred to as AIMD hereafter) with trajectory surface–hopping nonadiabatic molecular dynamics (NAMD) to survey CO-\ce{N2O} mixtures, compressed from ambient pressures to 160~GPa at 300–1500~K. 

All ground-state and finite-temperature calculations were performed with \textsc{VASP} \cite{kresse1993ab,kresse1996efficiency,kresse1996efficient}, using projector augmented wave (PAW) pseudopotentials \cite{blochl1994projector} and the PBE-D3 generalized gradient approximation (GGA) exchange-correlation functional \cite{perdew1996generalized,perdew1992atoms,becke1992density}. A plane-wave cutoff of \(850\;\mathrm{eV}\) and a single \((\tfrac14,\tfrac14,\tfrac14)\) \emph{k} point \cite{baldereschi1973mean} sufficed for 90- and 300-atom cubic supercells. While compressing along isotherms reducing box length by 1\% each step, each such (\(\rho,T\)) state was equilibrated for 5–10 ps using Born–Oppenheimer \textit{NVT} AIMD with a Nosé–Hoover thermostat fixing temperatures along the isotherm, scalar nonadiabatic couplings were used to gauge the necessity of NAMD simulations at each \(\rho,T\) point using an in-house code \cite{Chu2021JPCL}, and if determined to be necessary, 1-1.3 ps of NAMD using the the fewest switches surface hopping (FSSH) algorithm \cite{tully1990molecular} and directional non-adiabatic couplings were performed, using the same in-house code to couple VASP trajectories with nonadiabatic adjustments \cite{Akimov2013JCTC}. We term this the \emph{diabatic trajectory stitching} approach.

Much of modern high-pressure or combined-high pressure-temperature predictions for pure materials or mixtures involve AIMD, which implicitly implies Born-Oppenheimer dynamics. However, as already mentioned, analysis of CO and \ce{N2O} systems which involve unimolecular dissociation of nitroxides, via a singlet-to-triplet transition of detaching oxygen atoms as it forms \ce{O2} molecules, potentially requires nonadiabatic treatment. In addition, pressure itself acts as a major source of excitation, as shown in Fig. 1(a), in terms of decrease in the HOMO-LUMO gap in \textit{P-T} space. As can be seen, the excitation can reach up to 3-5 eV for CO-\ce{N2O}, which coupled with the downshift in electronic energy to a new PES upon dissociation can contribute to high levels of nonadiabaticity. In addition, the excited states population fraction $\eta(P, t)=\sum_{J>0} p_J(P, t)$ (see SI section I for description) also shows a significant rise with combined high–$P$/$T$ effects. 

It can be estimated from classical RRKM theory \cite{magee1952theory} how much ensemble time is required for such systems to have sufficient number of gas-phase collisions to enter a reactive state and form a product completely devoid of gaseous components (See SI section II for detailed discussion). For 90-300 atom cells, $\sim$45~ps of combined AIMD+NAMD is sufficient for 300-1500~K simulations. We distribute the NAMD time over 11-14 \(\rho,T\) points (dependent on isochore), ranging from 1-1.3 ps at each point. The rationale behind this fractional nonadiabatic treatment is the following: ionic configurational adjustments being stochastic require approximately one order of magnitude higher time to equilibrate \cite{curchod2018ab} than electronic state adjustments. In essence, we present a new paradigm for analyzing systems which are metastable and susceptible to nonadiabatic electronic transitions. It presents the community with a workflow that can be used to make predictions and match experiments while being physically consistent, something a simple AIMD may not suffice for. As shown in Fig. 1(b) by gray bars, in a typical 1200~K isothermal trajectory for \ce{CO}-\ce{N2O}, approximately 30.4\% of the total trajectory time is used to evolve the system nonadiabatically, mostly in regions corresponding to certain transition points. When the system collapses from molecular gas to a networked solid, we can track $ \chi(P)=\eta(P,t) \gamma_e(P) $ because it couples the amount of excited population to how strongly that population shifts with pressure. Here, the electronic Grüneisen parameter $ \gamma_e(P) $ is a dimensionless measure of volume-sensitivity of an excited-state energy \cite{YuCardona2010}. In \ce{CO}+\ce{N2O}, $\chi$ rises from $<0.3$ below 15 GPa to $\approx 2.2$ at the molecular-amorphous transition pressure, providing signal for impending structural collapse.

\begin{figure}[t]
\includegraphics[width=0.45\textwidth]{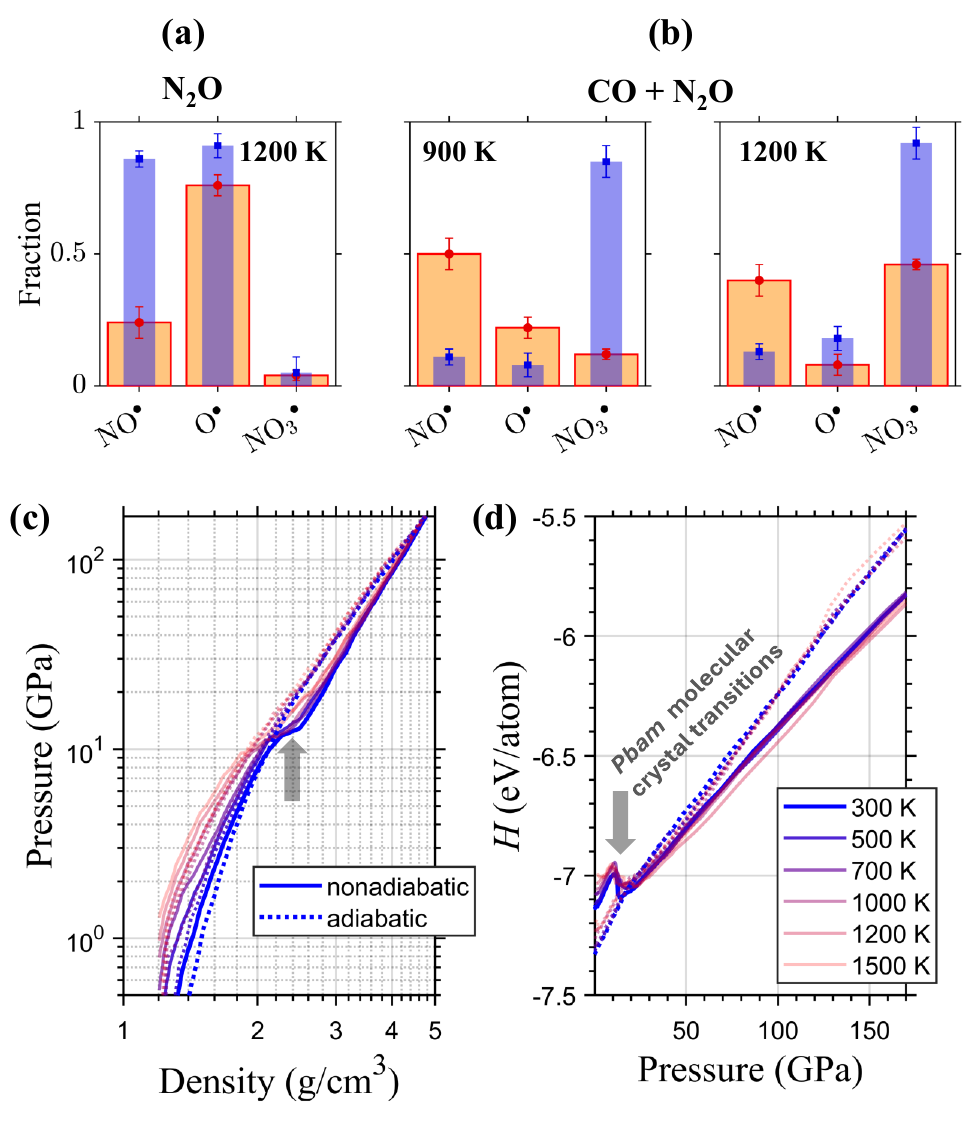} 
\caption{Fragmentation distribution histograms for \ce{NO^{$\bullet$}}, \ce{O^{$\bullet$}}, and \ce{NO3^{$\bullet$}} free radicals in (a) \ce{N2O} and (b) \ce{CO-N2O} at $\sim$20 GPa. Black lined boxes show the fraction of N- and O-atoms in particular free radicals, whereas orange bars show the contribution of NAMD-origin free radicals to total. (c) Pressure-density equation-of-state family of curves for compression of \ce{CO2} + \ce{N2O} using AIMD versus NAMD simulations are shown for multiple temperatures along a heat-compress pathway. (b) Similar plots for enthalpy versus pressure are shown. In both cases, composite curves obtained from stitching together data from molecular gas, \textit{Pbam} molecular crystal, and amorphous phases are plotted.}
\label{fig:2}
\end{figure}

\begin{figure*}[t]
\includegraphics[width=0.95\textwidth]{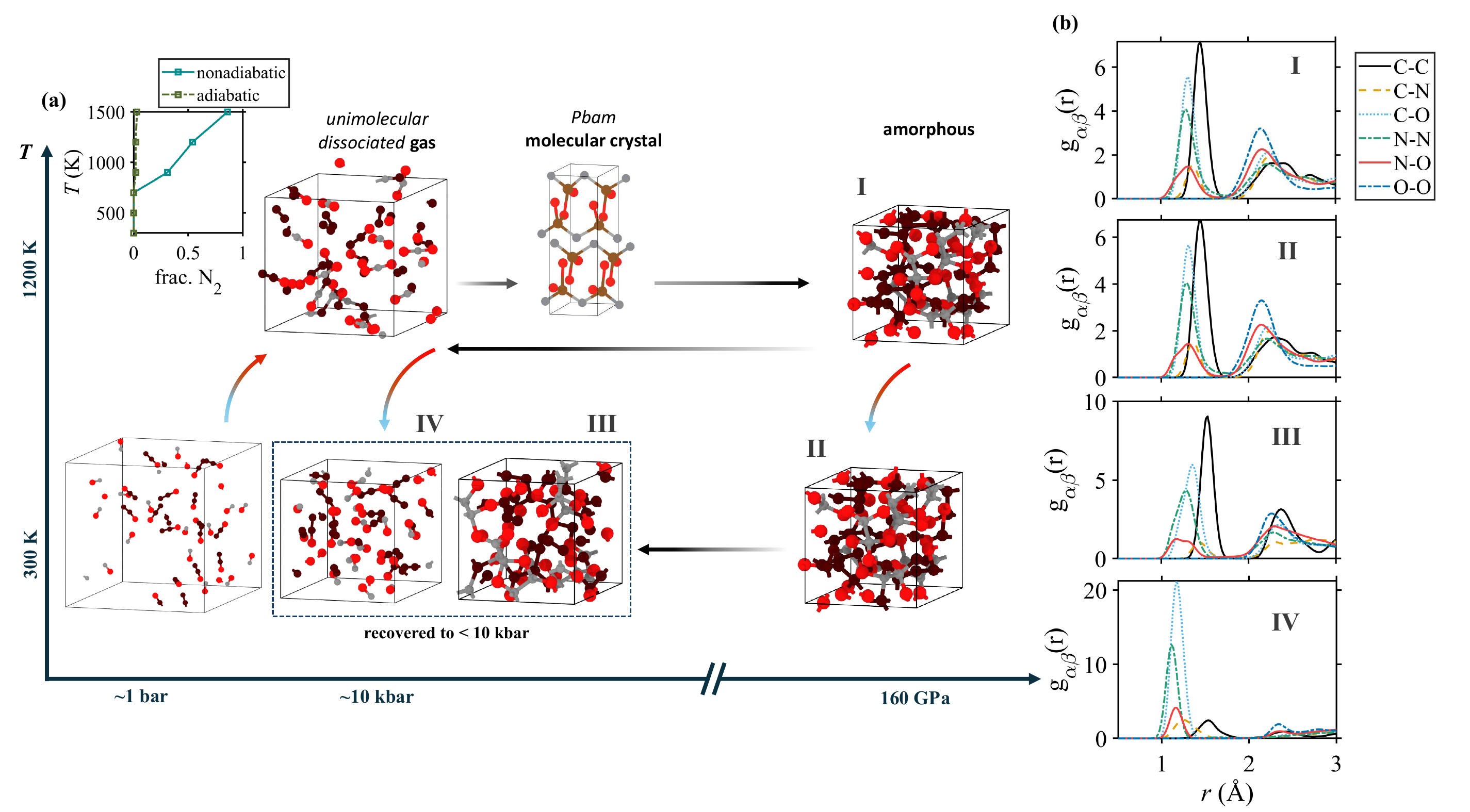} 
\caption{(a) Formation enthalpy of different carbon monoxide-nitroxide mixtures at 160 GPa, calculated w.r.t. 1 atm constituents, as a function of temperature for the heat-compress-quench path. (b) Formation enthalpy of different carbon monoxide-nitroxide mixtures at 10 kbar, calculated w.r.t. 1 atm constituents, as a function of temperature for the heat-compress-quench-decompress path. These thermodynamic pathways are shown in the top-right corner of both (a) and (b). (c) Snapshots of different structures resulting from starting with CO+\ce{N2O} molecular gas mixtures are shown in a pressure-temperature phase diagram. The radial distribution functions of four relevant structures are shown in right-column plots, while the fractionation of \ce{N2O} to \ce{N2} is shown as a function of temperature in the inset on top-left.}
\label{fig:3}
\end{figure*}

What this type of nonadiabatic refactoring describes is shown using the energy diagram in Fig. 1(c), which shows the sequence of hopping across at least three PESs: gaseous non-dissociated (S$_{non-d.c.}$) to dissociated (S$_{d.c.}$) at $<$4 GPa; $>$75\% fraction formation of molecular C-N-O crystal or polymeric amorphous C-N-O ($S_{poly.}$) phase at $\sim$23 GPa and $\sim$ 68 GPa respectively. Each such hopping involves a conical intersection across two PESs, and since such PESs are not continuously differentiable \cite{yarkony1996diabolical}, addressing the chemical evolution of this system using machine learned interatomic potentials might lead to fictitious results. This, furthermore, makes analysis of such C-N-O systems even more challenging.

Surface-hopping NAMD trajectories exhibit hopping probabilities of up to 15 \% per fs for O-transfer intermediates, with Landau–Zener coupling peaking near C–O = 1.85~Å. The associated non-radiative energy release heats the lattice locally by $\sim$150~K within 200~fs, accelerating further bond rearrangements — a feedback that is absent in strictly adiabatic AIMD. This reduces the onset of CO polymerization in the \ce{CO}-\ce{N2O} mixtures from $\sim$8~GPa \cite{Evans2006ChemMater} in case of pure \ce{CO} to $<$5~GPa in C-N-O mixtures. While this confirms one of our hypothesis, there wasn't much gain to be made anyway, in terms of reducing the polymerization-onset of CO, given the low onset pressure to begin with.  

For benchmarking, we used NAMD to compare with reported data on thermalized \ce{N2O} gas. The dissociation of \ce{N2O} ($\Sigma^+$) to \ce{N2} and \ce{O^{$\bullet$}}(${ }^3 P$) is considered to be principal case for such benchmarking, although data exists for \ce{NO}, \ce{NO2}, and \ce{N2O5} as well \cite{loirat1985thermal,papapolymerou1985unimolecular,karabeyoglu2008modeling,gill1958theoretical,johnston1986unimolecular}. This decomposition is expected to be complete at 1000-1200 K at $\sim$MPa pressures \cite{loirat1985thermal}. As shown in Fig. 2(a), $\sim$76\% breakdown of the \ce{N2O} molecules is achieved in the NAMD at 1200~K, with $>$91\% of the contribution arising from the NAMD trajectory. The dynamics is slightly altered when CO is present in the system. In that case, \ce{NO^{$\bullet$}} forms in the adiabatic trajectory unlike in the case of \ce{N2O} by itself. However, more consequential is the presence of $\sim$46\% \ce{NO3^{$\bullet$}} free radical at $\sim$1200~K, overwhelmingly ($\sim$92\%) originating from the NAMD trajectory. This species plays a key role in decrystallization and amorphization characteristics of C-N-O mixtures by virtue of its high reactivity \cite{karagulian2007heterogeneous}.

\begin{figure*}
\includegraphics[width=0.85\textwidth]{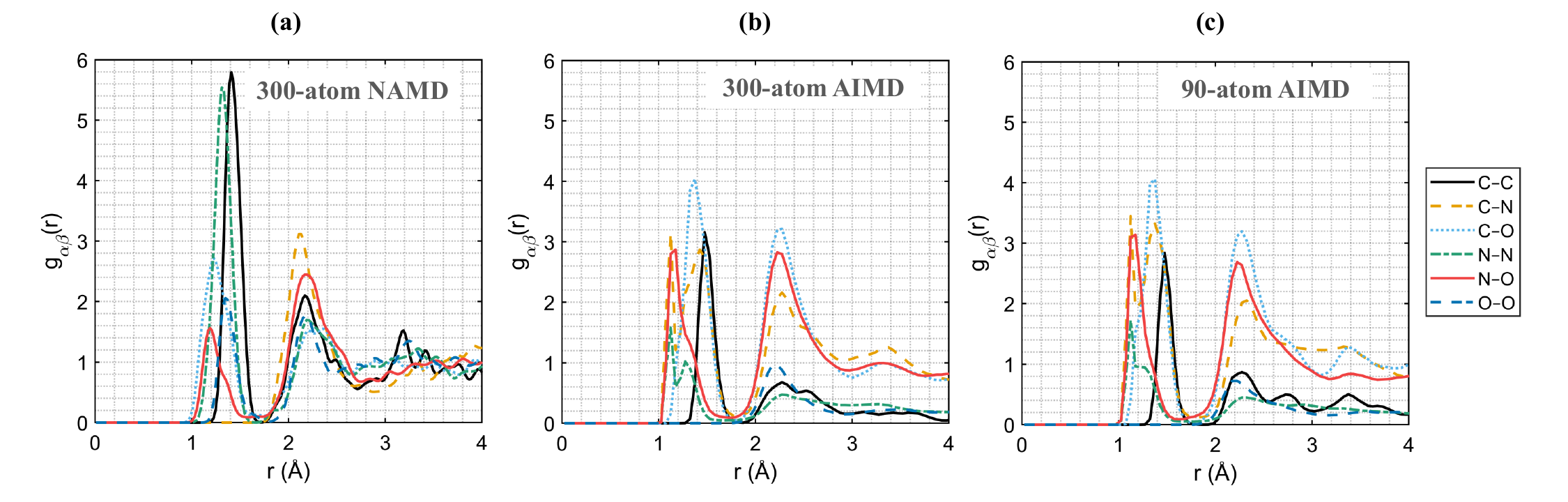} 
\caption{Radial distribution functions of C-N-O amorphous structures formed from compression of CO+\ce{N2O} mixture along the 1200~K isotherm using (a) 300-atom nonadiabatic molecular dynamics, (b) 300-atom Born-Oppenheimer AIMD, and 90-atom Born-Oppenheimer AIMD. Comparative analysis shows drastically different second shell characteristics, alongside stronger O-O bonds, and a higher fraction of N-N bonds. This is commensurate to a lower quantity of C-N and C-O bonds.}
\label{fig:4}
\end{figure*}

The effect of the presence of this free radical can be seen in terms of changes in bulk thermodynamic variables: Fig. 2(c) shows the equation of state for the adiabatic and nonadiabatic cases at multiple temperatures, whereas Fig. 2(d) shows the same for enthalpy versus pressure. In either case, a kink is observed in the family of curves for the nonadiabatic case, irrespective of temperature, that can be attributed to not just accurately capturing the dissociation characteristics of \ce{N2O} at high temperatures, but also the \ce{NO3^{$\bullet$}} free radical at all temperatures. This correction leads to a reduction in net enthalpy that can be obtained from the system. Thus, the downstream effects on added nonadiabaticity is not just altered and closer-to-reality chemical dynamics but also more accurate energetics.

At constant temperature the enthalpy is $ H(P,T)=U(P,T)+P\,V(P,T)$ so its pressure derivative is
$ \bigl(\partial H/\partial P\bigr)_T \;=\;V -T\,\bigl(\partial V/\partial T\bigr)_P $.
Three contributions control the sign of $H$: Mechanical work \(+P\,\mathrm{d}V\) $-$ always \emph{positive}, forcing \(H\) upward as the lattice/fluid is compressed. Thermal expansivity term \(-T(\partial V/\partial T)_P\) $-$ negative for fluids, nearly zero for rigid amorphous solids. Chemical‐bond term \(\Delta U_{\mathrm{chem}}\) $-$ exothermic (\(<0\)) when polymerization or network formation releases bond energy, endothermic (\(>0\)) if dissociation dominates. Below $\sim$15\,GPa the mixture is a van-der-Waals fluid or molecular crystal, with CO amorphization; enthalpy $H$ rises almost linearly with pressure $P$ because only the mechanical term $P\,V$ is important. Near $P\!\approx\!15$–20\,GPa a sudden and cooperative
\hbox{molecular-to-amorphous‐network} transition \emph{drops} $H$, over-compensating the $P\,V$ term for a narrow window. Above $\sim$30\,GPa the network starts becoming fully connected;
its smaller molar volume means \((\partial H/\partial P)_T\) is again positive, but with a gentler slope than in the molecular regime. If the transition were endothermic—e.g.\ simple dissociation into weaker bonds—no such dip would appear and $\Delta H(P)$ would increase monotonically.  Thus, from purely mechanistic/thermodynamic considerations, sigmoidal curves as in Fig.~2(c,d) nonadiabatic cases, are to be expected and these reflect a competition between \textit{mechanical compression} (always raising the enthalpy) and \textit{exothermic bond rearrangement} (lowering it over a limited pressure interval).


Let's examine the temperature–pressure ($T$–$P$) phase diagram of \ce{CO + N2O} mixtures for both high-pressure structures as well as for recoverability considerations. This is because synthesizing an energetic amorphous material is only a meaningful exercise if the said material can be recovered to near-ambient conditions. Beginning with a molecular gas at roughly $1\,\text{bar}$, heating the system induces unimolecular dissociation above about $900\,\text{K}$. The inset of Fig.~3c shows that this dissociation starts at $900\,\text{K}$ and is nearly complete by $1500\,\text{K}$. After dissociation, further compression produces a molecular crystal (space group \textit{Pbam}; at 25 GPa: $a = 7.22\,\text{\AA}$, $b = 3.25\,\text{\AA}$, $c = 1.971\,\text{\AA}$, $\alpha = 90^{\circ}$, $\beta = 90^{\circ}$, $\gamma = 90^{\circ}$), which, under continued compression, transforms into an amorphous structure. This molecular crystal, however, can only be obtained using constrained metadynamics. Once the compressed ($160\,\mathrm{GPa}$, $1200\,\mathrm{K}$) amorphous structure I is attained , there are two possible pathways to attempt to recover that structure back to near-ambient conditions. The first approach (structure IV) is to decompress along the $1200\,\mathrm{K}$ isotherm from $160\,\mathrm{GPa}$ down to approximately $10\,\mathrm{kbar}$, and then quench to $300\,\mathrm{K}$. The second approach (structure III) is to quench first at $160\,\mathrm{GPa}$ from $1200\,\mathrm{K}$ to $300\,\mathrm{K}$ and subsequently decompress from $160\,\mathrm{GPa}$ to $10\,\mathrm{kbar}$. As shown by the radial distribution functions in Fig. 3(c), the recovered structure III retains its amorphous network, whereas structure IV begins to break down into molecular constituents. We select 10 kbar as the recovery pressure because fluctuations in MD might otherwise take the system sporadically to negative pressures, which adversely affects making any physically-relevant conclusions. The recovery/decompression trajectory was simulated by increasing box length by 1\% at each \(\rho,T\) point, until ensemble-averaged pressure dropped below 10~kbar. 22.6-24~ps of trajectory was simulated, with 3.6-5.2~ps using NAMD simulations, utilizing the same \emph{diabatic trajectory stitching} approach as discussed earlier. It is very important to note here that the decompressed amorphous phase IV does not break down to molecular components when nonadiabaticity is not used in the trajectory.

It is imperative to understand the feature differences between the compressed-reactive product we get from NAMD and AIMD simulations. We pick the edge case of \textit{P} $\approx$ 160~GPa and at \textit{T} $\approx$ 1200~K, and compare the features in the radial distribution function as shown in Fig. 4. Nonadiabaticity promotes enhanced dissociation of \ce{N2O} and alternate pathways to polymerization which is represented by structural differences, i.e. reduced fractions of C-N and C-O bonds, stronger O-O bonds and higher fraction of N-N bonds. The accurate simulation of \ce{N2O} disproportionation introduces highly reactive NO and O free radicals which assists in enhanced formation of O-O and N-N bonds.

This letter underscores the need for combining nonadiabatic calculations with Born-Oppenheimer molecular dynamics to capture the complex phase behavior of reactive C-N-O mixtures under pressure. We find that the dissociation and the nature of amorphization are driven by pressure-enhanced \ce{N2O} decomposition, with structural signatures imprinted in enthalpy and recoverability characteristics. Recoverability of high-pressure phases is sensitive to the path taken, emphasizing the role of local energy metrics in high-pressure physics and chemistry.


This work was performed under the auspices of the U.S. Department of Energy by Lawrence Livermore National Laboratory (LLNL) under contract number DEAC52-07NA27344. The authors acknowledge funding support from the DOE Laboratory Directed Research and Development (LDRD) program at LLNL under the project tracking code 23-ER-028. We would like to thank Livermore Computing (LC) for helping us in finishing this extremely expensive study by utilizing $\sim$28 million cpu-hours. 


\bibliography{refs.bib}
\bibliographystyle{apsrev4-1}

\newpage


\clearpage

\section{SUPPLEMENTARY INFORMATION}

\section{I. Excited states population fraction}

$$
\eta(P, t)=\sum_{J>0} p_J(P, t) .
$$

Here, $\eta(P, t)$ is the excited-state population fraction (dimensionless, $0 \leq \eta \leq 1$ ) at external parameter $P$ (in this context, pressure) and time $t$. $J$ is the index of adiabatic electronic states ordered by energy; $J=0$ is the ground state, $J>0$ are excited states. The sum runs over all excited states.
$p_J(P, t)$ is the population (probability) of being on electronic state $J$ at $(P, t)$. 

In practice, when using FSSH/NA-MD based on electronic coefficients, $\left.P_J=\left.\langle | c_J(t)\right|^2\right\rangle_{\text {traj }}$ and when active-surface occupancy is used: $P_J=\frac{1}{N_{\text {traj }}} \sum_k \mathbf{1}\left[J_k^{\text {active }}(t)=J\right]$.

Normalization: $\sum_{J \geq 0} P_J(P, t)=1$, so equivalently $\eta(P, t)=1-P_0(P, t)$.

Notes: If states have degeneracy $\boldsymbol{g}_{\boldsymbol{J}}$, either treat each component as a separate $J$ in the sum or weight by $g_J$. If the basis is truncated, ensure populations are renormalized so the total remains 1 .

\section{II. RRKM/NA-TST times of first-event times for CO + N\texorpdfstring{$_2$}{2}O mixtures}

We estimate the \emph{waiting time to the first reactive event} that seeds amorphization during isothermal compression of mixed CO + N$_2$O, using RRKM/statistical kinetics augmented by nonadiabatic transition-state theory (NA-TST). The objective is a predictive, box-size-dependent criterion showing what time limit of \textit{ab initio}/nonadiabatic molecular dynamics is \emph{sufficient} to observe the onset of chemistry for both \textbf{90-atom} and \textbf{300-atom} cells over \SIrange{300}{1500}{K}. We provide closed-form scaling, a Landau–Zener nonadiabatic factor, and tabulated median ($t_{50}$) and conservative ($t_{95}$) first-event times.

\subsection{Calculate Reactive Rate Constants for Gas-amorphous transition}
\subsection{Per-encounter \textit{gas-like} capture Model}
We adopt a capture–complex–branching picture for short-range CO$\cdots$N$_2$O encounters. The effective \emph{bimolecular} reactive rate constant is
\begin{equation}
k_2(T)=\langle \sigma v\rangle_T \; P_{\mathrm{react}}(T,P),
\label{eq:k2}
\end{equation}
where $\langle \sigma v\rangle_T$ is the kinetic capture factor and $P_{\mathrm{react}}$ is the \emph{per-encounter} success probability (RRKM branching within the energized complex multiplied by a nonadiabatic hop probability). For hard-sphere kinetics
\begin{equation}
\sigma=\pi\!\left(\frac{d_{\mathrm{CO}}+d_{\mathrm{N_2O}}}{2}\right)^2,\quad 
\end{equation}
\begin{equation}
\langle v\rangle=\sqrt{\frac{8k_{\mathrm B}T}{\pi\mu}},\quad \text{and}
\end{equation}
\begin{equation}
\mu=\frac{m_{\mathrm{CO}}m_{\mathrm{N_2O}}}{m_{\mathrm{CO}}+m_{\mathrm{N_2O}}}.
\label{eq:sigv}
\end{equation}
We use kinetic diameters $d_{\mathrm{CO}}=\SI{3.76}{\angstrom}$ and $d_{\mathrm{N_2O}}=\SI{3.30}{\angstrom}$ \cite{Breck1974,Matteucci2006}.

For the energized collision complex, the RRKM microcanonical branching ratio for product formation is
\begin{equation}\label{eq:rrkm}
\begin{split}
p_{\mathrm{prod}}(E)
&= \frac{k_{\mathrm{prod}}(E)}{k_{\mathrm{prod}}(E)+k_{\mathrm{back}}(E)}\\
&= \frac{N^{\ddagger}_{\mathrm{prod}}(E-E_0)}{N^{\ddagger}_{\mathrm{prod}}(E-E_0)+N^{\ddagger}_{\mathrm{back}}(E-E_0)}
\end{split}
\end{equation}
with standard densities/counts of states. A necessary hop at a diabatic crossing has the Landau–Zener probability\cite{Wittig2005}
\begin{equation}
P_{\mathrm{LZ}}(E)=1-\exp\!\left[-\frac{2\pi H_{ab}^2}{\hbar\, v_\perp(E)\,|\Delta F|}\right],
\label{eq:lz}
\end{equation}
where $H_{ab}$ is the electronic coupling, $\Delta F$ the diabatic slope difference, and $v_\perp$ the normal velocity through the crossing. Canonically averaged,
\begin{equation}
P_{\mathrm{react}}(T,P)=\big\langle p_{\mathrm{prod}}(E)\,P_{\mathrm{LZ}}(E)\big\rangle_T,
\end{equation}
which is the NA-TST extension of RRKM for spin-crossing or electronically nonadiabatic channels \cite{LykhinVarganov2016}.

\paragraph*{Waiting-time statistics in a periodic box.}
For equimolar CO and N$_2$O counts, $N_A=N_B$ in a box of volume $V$, the Poisson propensity for the \emph{first} A+B reaction is
\begin{equation}
\lambda = k_2(T)\,\frac{N_A N_B}{V},
\end{equation}
which yields exponential order statistics: $t_{50}=\ln 2/\lambda$ and $t_{95}=\ln 20/\lambda$.

\paragraph*{Useful scaling.}
At fixed $k_2$, $t\propto V/(N_A N_B)$; at fixed number density (i.e.\ $V\propto N_A+N_B$ for equimolar boxes), $t\propto 1/N$; and at fixed composition $t\propto 1/(\sigma \sqrt{T}\,P_{\mathrm{react}})$ via Eq.~\eqref{eq:sigv}. These relations are used below to compare 90- and 300-atom boxes and to rescale along isotherms during compression.

\subsubsection{Per-attempt \textit{solid-like} Eyring Model}
In highly compressed, nearly amorphous environments, ``collisions'' blur into repeated barrier-crossing \emph{attempts} with Eyring frequency $\nu=k_{\mathrm B}T/h$. The per-site rate is $k_{\text{site}}=P_{\mathrm{attempt}}\nu$, and the box-level first-event time for $N_{\mathrm{mol}}$ statistically independent sites is
\begin{equation}
t_p=\frac{-\ln(1-p)}{N_{\mathrm{mol}}\,P_{\mathrm{attempt}}\,k_{\mathrm B}T/h}.
\label{eq:attempt}
\end{equation}
This variant leads to the same linear-in-$N$ acceleration as the previous model and is convenient when $P_{\mathrm{attempt}}$ is read off directly from nonadiabatic MD (NAMD).

\subsection{Parameters, geometry, and compression}
We consider equimolar boxes of:
\begin{itemize}
\item \textbf{90 atoms}: $N_{\mathrm{CO}}=18$, $N_{\mathrm{N_2O}}=18$ (\(N_{\mathrm{mol}}=36\)).
\item \textbf{300 atoms}: $N_{\mathrm{CO}}=60$, $N_{\mathrm{N_2O}}=60$ (\(N_{\mathrm{mol}}=120\)).
\end{itemize}
At ambient pressure, the \textbf{300-atom} box has side $L=\SI{21.4}{\angstrom}$ ($V=\SI{9.80e-27}{m^3}$). Along isotherms to \(\sim\)\SI{160}{GPa}, the box length decreases by \(\sim\)\SI{34}{\percent}, so $V/V_0\approx 0.287$. Times below are given at ambient volume and, in parentheses, multiplied by $0.287$ as a conservative compressed-volume proxy.

\subsection{Calibration to NAMD: checking for 45-ps sufficiency}
NAMD/AIMD show that \textbf{45~ps suffices} to observe the onset of chemistry even in the \textbf{90-atom} box. Two equivalent calibrations reconcile the statistical model with this:

\paragraph*{(A) Per-encounter calibration (Capture model).}
Solving $t_{50}(90\ \text{atoms},\,\SI{300}{K})=\SI{45}{ps}$ for Eq.~\eqref{eq:k2}–\eqref{eq:sigv} gives an effective
\[
P_{\mathrm{react}}\approx 6\times 10^{-4}\ \ (0.06\%).
\]
By linear-in-$N$ scaling, the \textbf{300-atom} box requires only $\sim\!1.76\times 10^{-4}$ (\(0.0176\%\)) for the same $t_{50}$ at \SI{300}{K}, and less at higher $T$.

\paragraph*{(B) Per-attempt calibration (Eyring picture).}
With $\nu=k_{\mathrm B}T/h$, Eq.~\eqref{eq:attempt} yields $P_{\mathrm{attempt}}\approx 6.8\times 10^{-5}$ at \SI{300}{K} (and $\approx 1.37\times 10^{-5}$ at \SI{1500}{K}) to achieve $t_{50}=\SI{45}{ps}$ for the 90-atom box. These values lie in the common $10^{-5}$–$10^{-4}$ band for condensed-phase ($>$ MPa), nonadiabatic channels \cite{Wittig2005,LykhinVarganov2016}.

Both are consistent with the well-established pressure-enabled reactivity and polymerization/disproportionation of CO at modest GPa even near room temperature, which accelerates further at higher $P$ and $T$ \cite{Evans2006,Evans2005OSTI}.

We adopt a single \emph{floor} value \(X\equiv 100\times P_{\mathrm{react}}=0.10\%\) (compatible with meV-scale Landau–Zener couplings and rare-event crossings per close approach of PESs). With this \(X\), both boxes comfortably meet the 45~ps criterion across \SIrange{300}{1500}{K}; compression of boxes only shortens the times.

\subsubsection*{Median first-event time $t_{50}$ (ps) at $X=0.10\%$}
Numbers outside/inside parentheses correspond to ambient volume and compressed volume ($\times 0.287$), respectively.
\begin{table}[h]
\caption{$t_{50}$ (ps) vs.\ $T$ at $X=0.10\%$.}
\label{tab:t50}
\begin{ruledtabular}
\begin{tabular}{rcc}
$T$ (K) & 90 atoms & 300 atoms \\
\colrule
300  & 26.37 \,(7.57)  & 7.91 \,(2.27) \\
500  & 20.43 \,(5.86)  & 6.13 \,(1.76) \\
700  & 17.26 \,(4.95)  & 5.18 \,(1.49) \\
900  & 15.23 \,(4.37)  & 4.57 \,(1.31) \\
1200 & 13.19 \,(3.79)  & 3.96 \,(1.14) \\
1500 & 11.79 \,(3.39)  & 3.54 \,(1.02) \\
\end{tabular}
\end{ruledtabular}
\end{table}

\subsubsection*{Conservative $t_{95}$ (ps) at $X=0.10\%$}
We use $t_{95}=4.322\,t_{50}$ from exponential order statistics.
\begin{table}[h]
\caption{$t_{95}$ (ps) vs.\ $T$ at $X=0.10\%$.}
\label{tab:t95}
\begin{ruledtabular}
\begin{tabular}{rcc}
$T$ (K) & 90 atoms & 300 atoms \\
\colrule
300  & 113.98 \,(32.71) & 34.19 \,(9.81) \\
500  & 88.29 \,(25.34)  & 26.49 \,(7.60) \\
700  & 74.62 \,(21.42)  & 22.38 \,(6.42) \\
900  & 65.84 \,(18.90)  & 19.74 \,(5.66) \\
1200 & 56.95 \,(16.35)  & 17.09 \,(4.91) \\
1500 & 50.99 \,(14.64)  & 15.29 \,(4.39) \\
\end{tabular}
\end{ruledtabular}
\end{table}

\paragraph*{Interpretation.}
At ambient density, $t_{50}$ is $\ll \SI{45}{ps}$ for \emph{both} boxes across \SIrange{300}{1500}{K}; under compression ($V/V_0\!\approx\!0.287$), even $t_{95}$ falls below \SI{45}{ps}. If one wishes to reproduce exactly $t_{50}\!\approx\!\SI{45}{ps}$ at \SI{300}{K} in the \emph{90-atom} box at ambient density, setting \(X=0.06\%\) achieves this; the same $X$ gives $t_{50}\!\approx\!\SI{12.6}{ps}$ and $t_{95}\!\approx\!\SI{54.6}{ps}$ upon compression, again making \SI{45}{ps} a safe window.

\begingroup
\renewcommand{\thefigure}{S\arabic{figure}}
\setcounter{figure}{0} 
\begin{figure*}[hbtp]
  \centering
  \includegraphics[width=0.9\textwidth]{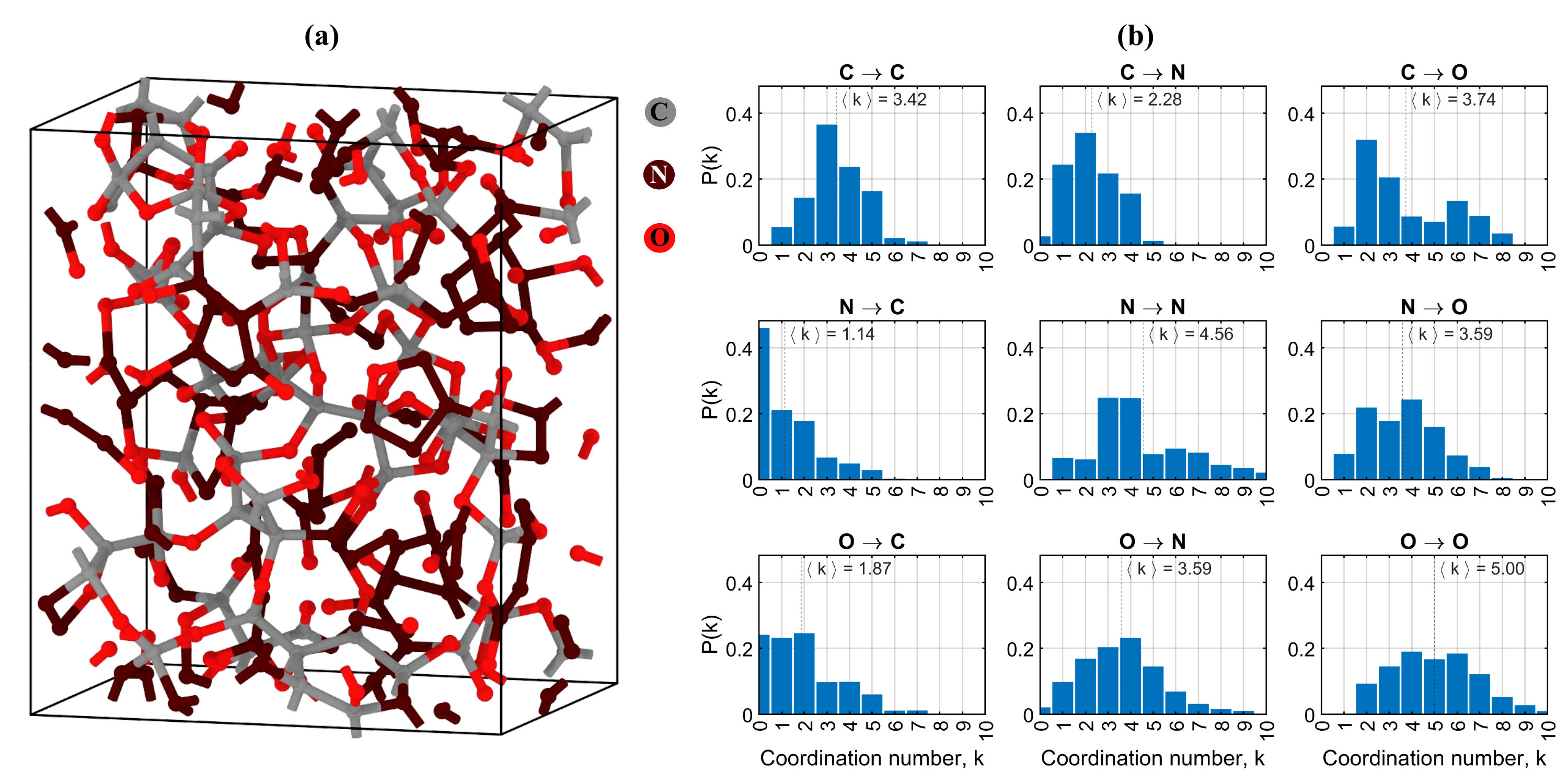} 
  \caption{(a) 300-atom final snapshot structure of amorphous C-N-O product, obtained at ~160~GPa and ~1200~K, and (b) nearest neighbor coordination number (NNCN) histograms of this configuration.}
  \label{fig:S1}
\end{figure*}
\endgroup

\subsection{Dilute gas-phase baseline }
For dilute-gas (ambient pressure), single-pulse shock-tube and infrared-emission studies of the \emph{direct} exchange channel CO + N$_2$O $\to$ CO$_2$ + N$_2$ over \(\sim\)\SIrange{1100}{2300}{K} report Arrhenius rate constants that, when compared to $\langle\sigma v\rangle$, imply per-encounter probabilities $\ll 10^{-3}\%$ even near \SI{1500}{K} \cite{LinBauer1969,FujiiCPL1985,LoiratJPC1987}. Such values are far too small to explain picosecond-scale initiation in dense, compressed cells. By contrast, carbon monoxide is known to react/polymerize at ps-scales readily above a few GPa (even near room temperature) in the solid/condensed environment,\cite{Evans2006,Evans2005OSTI} justifying an \emph{effective} $X$ in the $10^{-4}$--$10^{-3}$ range (and higher as compression proceeds), consistent with the values above. Thus, a 45-ps NAMD/AIMD might overpredict the transition pressure of amorphization because of inadequate simulations frames in the dilute-gas limit but at $>$MPa pressures, it should be accurate.

\subsection{Conclusions}
For any alternative stoichiometry or density path, we use
\[
t_{50}=\frac{\ln 2}{k_2}\,\frac{V}{N_A N_B},\qquad
k_2=\langle\sigma v\rangle\,X/100,
\]
with Eqs.~\eqref{eq:sigv} . Time scales linearly with $V$ at fixed $X$ and $T$, and inversely with $N_A N_B$.

With a floor \(X=0.10\%\), both \textbf{90-atom} and \textbf{300-atom} CO+N$_2$O boxes exhibit median first-event times well below \SI{45}{ps} across \SIrange{300}{1500}{K}, and $t_{95}<\SI{45}{ps}$ upon compression to \(\sim\)\SI{160}{GPa}. 



\section{III. General domain of pressure-induced dissociation and polymerization}

\subsection{Breakdown of the Born--Oppenheimer picture under compression}

We assess where nonadiabatic molecular dynamics (NA-MD) is required along the CO + N$_2$O compression path (0–200 GPa) using two yardsticks: the Landau--Zener adiabaticity parameter $\gamma$ and the electron--phonon thermalization time $\tau_{e\text{--}ph}$ relative to a characteristic nuclear period $\tau_n$.

\begin{equation}
\label{eq:LZ}
\gamma
= \frac{\hbar}{E_{eg}^2}
\left|
\dot{\mathbf R}\cdot
\left\langle
\Psi_e \left|
\frac{\partial H}{\partial \mathbf R}
\right| \Psi_g
\right\rangle
\right| ,
\end{equation}
where $E_{eg}$ is the instantaneous adiabatic energy gap between the ground ($\Psi_g$) and an excited ($\Psi_e$) electronic state, $\mathbf R$ are nuclear coordinates, and $\dot{\mathbf R}$ their velocities.

\noindent\textit{Rule of thumb $-$} Consider also the ratio of the electronic thermalization time $\tau_{e\text{--}ph}$ to a characteristic nuclear period $\tau_n$ ( $\approx 100~\mathrm{fs}$ for the high-frequency CO stretch). When either $\gamma \gtrsim 0.1$ or $\tau_{e\text{--}ph} \lesssim \tau_n$, you must step beyond ground-state Born--Oppenheimer MD.

\begingroup
\setcounter{table}{0}
\renewcommand{\thetable}{S\arabic{table}}
\begin{table*}[!hbtp]
\caption{Pressure-resolved guidance for when nonadiabatic MD is needed during CO + N$_2$O compression.}
\setlength{\tabcolsep}{4pt}
\renewcommand{\arraystretch}{1.15}
\footnotesize
\begin{tabularx}{\textwidth}{@{} C{1.6cm} L C{1.2cm} C{1.8cm} C{2.2cm} L @{}}
\hline\hline
\multicolumn{1}{l}{P range (GPa)} &
\multicolumn{1}{l}{Dominant physics} &
\multicolumn{1}{c}{$E_g$ (eV)} &
\multicolumn{1}{c}{$\tau_{e\text{--}ph}$ (fs)} &
\multicolumn{1}{c}{Need NAMD?} &
\multicolumn{1}{l}{Suggested method} \\
\hline
0$-$4 &
Cold molecular fluid &
7$-$8 &
$>5000$ &
No (BO okay) &
BO-AIMD / reactive FF \\
4$-$15 &
Spin-forbidden N$_2$O cleavage $\to$ CO$_2$ + N$_2$ (local crossings) &
6$-$7 &
3000$-$5000 &
Yes (local) at crossings &
Surface hopping $+$ SOC (ps windows) \\
15$-$20 &
Exothermic onset; first poly-CO / poly-N$_2$O links &
4$-$6 &
1000$-$2000 &
Mostly BO; Nonadiabaticity for chain-growth events &
SH/SHARC on reactive trajectories \\
20$-$65 &
C$-$N$-$O framework growth; shrinking gap &
2$-$3 &
300$-$800 &
Borderline (Nonadiabaticity for energy-flow studies) &
Ehrenfest/SHARC on subsets \\
65$-$100 &
Insulator $\to$ semiconductor crossover &
0.5$-$2 &
50$-$300 &
Growing nonadiabaticity (electrons store $P\,dV$ work) &
Ehrenfest-TDDFT / two-temperature AIMD \\
$\sim$90 &
Band-gap collapse / incipient metallization &
$\lesssim 0.4$ &
10$-$50 &
System-wide BO breakdown &
Decoherence-corrected Ehrenfest / NEAIMD \\
100$-$160 &
Metallic C$-$N$-$O alloy (degenerate electrons) &
0 &
5$-$20 &
NAMD mandatory &
Two-temperature NA-AIMD / MDEF-TDDFT \\
\hline\hline
\end{tabularx}
\end{table*}
\addtocounter{table}{-1}
\endgroup

\noindent In practice, BO descriptions remain reliable for equilibrium structure and coarse thermodynamics up to $\sim$60 GPa, but specific events (spin crossover; conical-intersection bond exchanges) already require nonadiabatic treatment in the 6$-$30 GPa window. Once the band gap collapses near $\sim$90 GPa, electrons cannot follow the nuclei adiabatically anywhere in the cell, and fully nonadiabatic dynamics become necessary through the 100$-$200 GPa metallic regime.

\section{IV. Trajectory in \textit{NVE} NAMD/AIMD simulations}

\begingroup
\renewcommand{\thefigure}{S\arabic{figure}}
\setcounter{figure}{1} 
\begin{figure*}[t]
  \centering
  \includegraphics[width=0.9\textwidth]{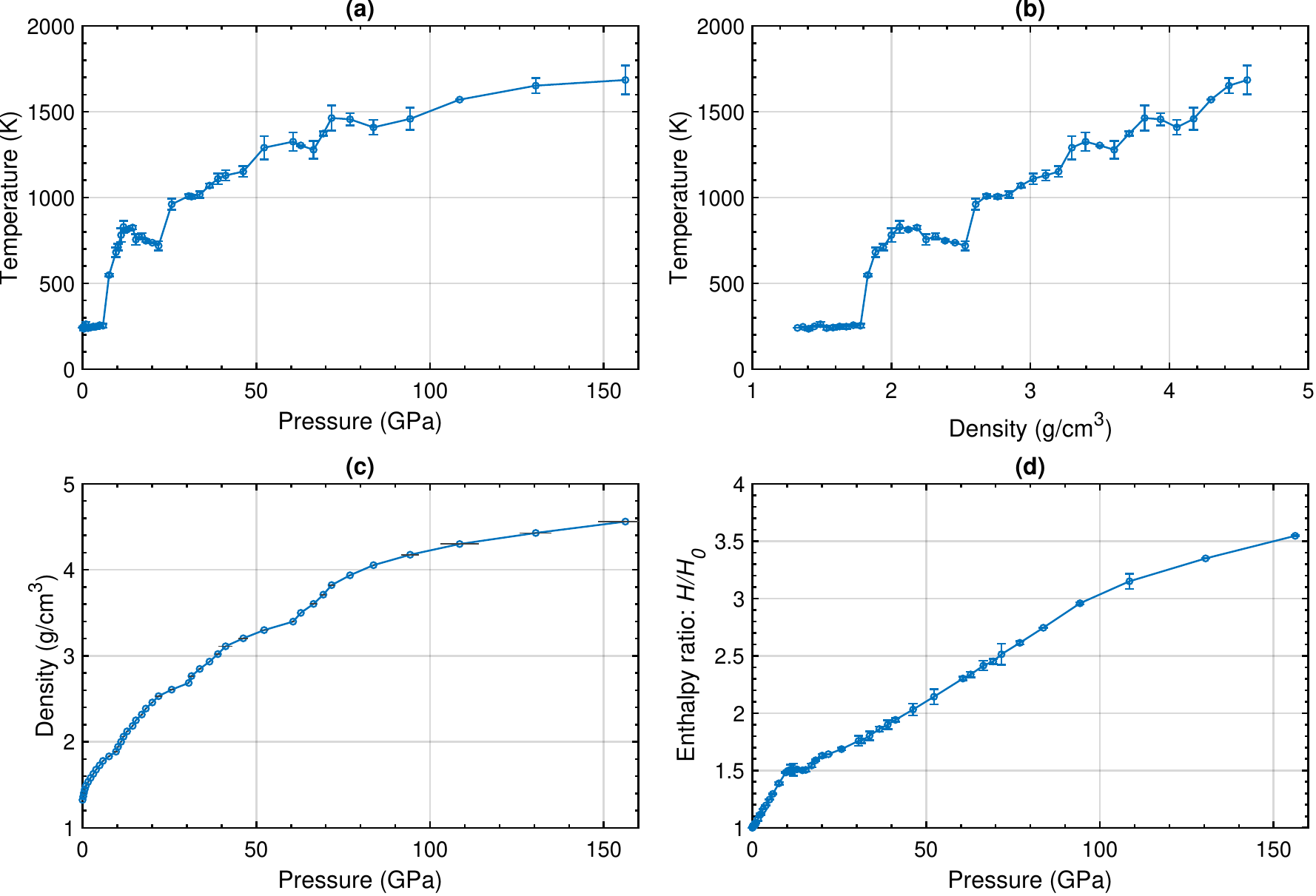} 
  \caption{Equation of state data from \textit{NVE} trajectory simulations: (a) temperature versus pressure, (b) temperature versus density, (c) density versus pressure, and (d) enthalpy ratio relative to ambient enthalpy versus pressure.}
  \label{fig:S2}
\end{figure*}
\endgroup

Starting from the initial 300-atom configuration at ambient pressure and room temperature, each \textit{NVE} point is obtained by a geometric volume update (1\% side length reduction each step) enforcing unique densities and a final $\rho=\SI{4.56}{g\,cm^{-3}}$ at \SI{156.2}{GPa} over a simulation duration of $\sim$52.0~ps ($\sim$8.8~ps in NAMD). Thermal evolution is adiabatic within piecewise regimes, with small latent releases near transitions; a localized thermal spike is seen in the 10$-$20~GPa window indicative of exothermic reaction. We report $H/H_0=(U+PV)/H_0$ (with $H_0$ the ambient enthalpy), constrained to start at 1 and rises to 3.56 at 156.2~GPa.

\emph{Regime I:} 0$-$4~GPa (condensed gaseous/molecular). $T(P)$ nearly flat; $\rho(P)$ increases smoothly; $H/H_0$ rises rapidly from 1 due to $PV$ work with minimal chemistry. \\
\emph{Regime II:} 4$-$15~GPa (onset of chemistry). Gentle heating; first kink in $T(P)$ and $H/H_0(P)$; $\rho(P)$ remains monotone. \\
\emph{Regime III:} 15$-$20~GPa (polymerization/lock-in). Signature triad is observed: (i) a \emph{dip} in $H/H_0(P)$ indicating net stabilization, (ii) a \emph{plateau} (very flat to slightly negative slope) in $\rho(P)$ over 15.2$-$19.6~GPa, and (iii) a \emph{local spike} in $T(P)$ consistent with exothermic bond reconfiguration.  \\
\emph{Regime IV:} 20$-$65~GPa (framework growth). $H/H_0$ resumes a positive slope but smaller than pre-dip; $T(P)$ increases with reduced slope as the network thickens; $\rho(P)$ recovers to a smooth rise. \\
\emph{Regime V:} 65$-$100~GPa (gap-shrinking/ionic contribution). Continued densification; modest kinks in $T(P)$ and $H/H_0(P)$. \\
\emph{Regime VI:} 100$-$156~GPa (deep densification). Smooth approach to the final constraints ($\rho=\SI{4.56}{g\,cm^{-3}}$, $T=\SI{1685}{K}$, $H/H_0\simeq3.56$).

As can be seen in Fig. S1(a), $T(P)$ rises stepwise with a localized spike in the \SIrange{10}{20}{GPa} window, signaling the onset of network-forming chemistry. 
Fig. S1(b) shows $T$ against $\rho$ recasts the same thermal excursion at $\rho\!\approx\!\SIrange{1.8}{2.1}{g\,cm^{-3}}$ while preserving the strictly monotonic compression path. 
Fig. S1(c) shows the mechanical response, $\rho(P)$, exhibits a pronounced plateau—with near-zero to slightly negative slope—between \SI{10.2}{GPa} and \SI{19.6}{GPa}, followed by a smooth recovery into a stiffer densification regime above \SI{20}{GPa}. 
Fig. S1(d) shows the energetics, $H/H_{0}(P)$ shows a shallow dip over \SIrange{10}{20}{GPa} and then a gradual rebound to $\approx\!3.56$ by \SI{156}{GPa}, with a reduced post-event slope relative to the pre-event segment. 
Taken together, Fig. S1(a)–(d) trace a coherent sequence: molecular condensation $\rightarrow$ exothermic, nonadiabatic bond reconfiguration (dip/plateau/spike triad) $\rightarrow$ framework growth and progressive densification.

\newpage
\thispagestyle{empty}

\end{document}